\documentclass[11pt,a4paper]{article}
\usepackage{jcappub}
\usepackage{bm}
\usepackage{amsmath}
\def\dd{\mathrm{d}}

\def\mcP{\mathcal{P}}
\def\mcR{\mathcal{R}}

\def\Mpl{M_{\rm Pl}}

\def\0{{(0)}}
\def\sig0{\dot{\sigma}_0}
\def\dsig{\delta \sigma}
\def\dsigd{\dot{\delta\sigma}}

\def\dphi{\delta \phi}
\def\ph0{\dot{\phi}_0}
\def\dphid{\dot{\delta \phi}}

\def\dn{\Delta n}

\title{
Statistically Anisotropic Tensor Modes from Inflation
}
\author[a]{Tomohiro Fujita,}
\author[a]{Ippei Obata,}
\author[a,b]{Takahiro Tanaka}
\author[c,d]{and Shuichiro Yokoyama}
\affiliation[a]{Department of Physics, Kyoto University, Kyoto, 606-8502, Japan}
\affiliation[b]{Center for Gravitational Physics, Yukawa Institute for Theoretical Physics, Kyoto University, Kyoto 606-8502, Japan}
\affiliation[c]{Department of Physics, Rikkyo University, 3-34-1 Nishi-Ikebukuro, Toshima, Tokyo 171- 8501, Japan}
\affiliation[d]{Kavli IPMU (WPI), UTIAS, The University of Tokyo,
Kashiwa, Chiba 277-8583, Japan}
\emailAdd{t.fujita@tap.scphys.kyoto-u.ac.jp}
\emailAdd{obata@tap.scphys.kyoto-u.ac.jp}
\emailAdd{t.tanaka@tap.scphys.kyoto-u.ac.jp}
\emailAdd{shuichiro@rikkyo.ac.jp}
\abstract{
We consider the inflationary universe with a spectator scalar field coupled to a $U(1)$ gauge field and calculate curvature perturbation and gravitational waves (GWs). We find that the sourced GWs can be larger than the one from vacuum fluctuation and they are statistically anisotropic as well as linearly polarized. The GW power spectrum acquires higher multipole moments as $\mathcal{P}_h\propto (1-\cos^2\theta+\cos^4\theta-\cos^6\theta)$
irrespective of the model parameters.}

\keywords{inflation, primordial gravitational waves}
\arxivnumber{XXXX.XXXXX}

\usepackage{amsfonts}

\begin{document}

\begin{flushright}
KUNS-2713, RUP-18-1
\end{flushright}

\maketitle

%
%
%
\section{Introduction}
The recent detection of gravitational waves by the LIGO/VIRGO collaboration \cite{Abbott:2016blz} marked the beginning of gravitational wave astronomy.
The observation of  gravitational waves (GWs) from various astrophysical sources will give us novel opportunities to reveal outstanding problems
in astronomy and gravitational science. 
In addition, it is expected that primordial GWs originating from cosmic inflation in the early universe may be detected by future direct or indirect observations.  
For instance, LiteBIRD satellite~\cite{Matsumura:2013aja} and CMB-S4 project~\cite{Abazajian:2016yjj} aim to detect the B-mode polarization of the cosmic microwave background (CMB) which is produced by primordial GWs, while LISA~\cite{AmaroSeoane:2012km} and DECIGO~\cite{Kawamura:2006up} missions target its direct detection. 
Thus we have a good chance to probe fundamental physics through investigating primordial GWs. 
Under such circumstances, it is very important to explore a new possibility of generating GWs in the early universe.

The conventional inflationary scenario assumes a single-field slow-roll scalar field model in which the primordial GWs are produced from the vacuum fluctuation during inflation. Its power spectrum is generally characterized by the following properties: (i) its spectral shape is nearly scale-invariant, (ii) its amplitude is solely determined by the energy density of inflation, (iii) the two polarization modes of the GWs have the same amplitude, and (iv) it is statistically isotropic.
 However, several recent studies have shown that this picture is not necessarily realized, if we consider the other sources of GWs in the inflationary universe.
For example, 
in models where a gauge field is coupled to a scalar field or a pseudo-scalar field,
the perturbation of the gauge field can be amplified due to a tachyonic instability.
Intriguingly, the amplified gauge field perturbation sources scalar, vector and tensor fluctuations and can significantly enhance them.
Such kind of models have been extensively studied as a mechanism of generating non-Gaussianities observable in CMB~\cite{Linde:2012bt, Barnaby:2010vf, Barnaby:2011qe, Barnaby:2012tk, Anber:2012du, Barnaby:2012xt, Ferreira:2014zia, Peloso:2016gqs, Agrawal:2017awz}, present intergalactic magnetic fields~\cite{Barnaby:2012tk, Ratra:1991bn, Martin:2007ue, Demozzi:2009fu, Kanno:2009ei, Fujita:2012rb, Fujita:2013pgp, Fujita:2014sna, Obata:2014qba, Fujita:2015iga, Caprini:2017vnn},  primordial black holes~\cite{Linde:2012bt, Garcia-Bellido:2016dkw, Domcke:2017fix, Garcia-Bellido:2017aan}, baryon asymmetry \cite{Maleknejad:2014wsa, Fujita:2016igl, Caldwell:2017chz, Jimenez:2017cdr} and a sizable amount of primordial GWs \cite{Garcia-Bellido:2016dkw, Garcia-Bellido:2017aan, Anber:2012du, Barnaby:2012xt, Peloso:2016gqs, Sorbo:2011rz, Cook:2011hg, Dimastrogiovanni:2012ew, Adshead:2013qp, Mukohyama:2014gba, Obata:2014loa, Choi:2015wva, Namba:2015gja, Obata:2016tmo, Ito:2016aai, Domcke:2016bkh, Maleknejad:2016qjz, Guzzetti:2016mkm, Obata:2016xcr, Dimastrogiovanni:2016fuu, Adshead:2016omu, Obata:2016oym, Fujita:2017jwq, Thorne:2017jft}.
Remarkably, the properties of GW power spectrum listed above are altered, if the primordial GWs sourced by the gauge field acquire a relevant amplitude: (i) The GW power spectrum can be strongly scale-dependent, (ii) its amplitude would be no longer solely determined by the inflation energy scale, and (iii) the two polarization modes may have different amplitudes, which are totally different signatures from that of vacuum fluctuations.

In this paper, we explore the possibility of generating (iv) a testable statistical anisotropy of GW power spectrum sourced by a $U(1)$ gauge field which has a kinetic coupling to a scalar field during inflation.
Owing to the coupling, the kinetic energy of the scalar field is transferred to the gauge field, and the amplitude of the gauge field can grow on large scales. 
Hence, a background vector field naturally appears and breaks the isotropy of the universe.
Because of this broken rotational invariance, the fluctuation of vector field is coupled to scalar and tensor perturbations at linear level, and then provides the statistical anisotropies in their spectra.

 Historically, the generation of statistical anisotropy in the curvature perturbation has been discussed in the context of  the anisotropic inflation model~\cite{Watanabe:2009ct, Watanabe:2010fh, Kanno:2010nr, Watanabe:2010bu, Soda:2012zm, Bartolo:2012sd, Ohashi:2013qba, Ohashi:2013pca, Naruko:2014bxa, Ito:2015sxj, Abolhasani:2015cve, Ito:2017bnn}, motivated to explain the quadrupole anisotropy in the WMAP data reported by \cite{2010ApJ...722..452G}.
 However, current CMB observations restricts such an anisotropy to be smaller than  $\mathcal{O}(10^{-2})$~\cite{Kim:2013gka, Ade:2015lrj}, which implies that in the framework of anisotropic inflation the anisotropy of the GW power spectrum should be much smaller and therefore is difficult to be observed. Moreover, 
it was pointed out that
the attractor solution of the background dynamics is unavailable due to the stochastic effect
~\cite{Fujita:2017lfu}.
Recently, another inflationary scenario with higher spin particles has been developed which leaves an imprint of multipole moments higher than quadrupole in the two-point function of the curvature perturbation \cite{Kehagias:2017cym, Bartolo:2017sbu, Franciolini:2017ktv}. However, their effects on the GW power spectrum are yet to be explored. 

In this work, we consider the possibility that a $U(1)$ gauge field is coupled not to the inflaton but to a spectator scalar field to overcome the above shortcomings of the original anisotropic inflation model.
In this case, the generation of the statistically anisotropic curvature perturbation is suppressed and an attractor solution is available.
At the same time, interestingly, the amplified gauge field perturbations on super-horizon scales can source GWs with a sizable amount of statistical anisotropies in the GW power spectrum.
Intriguingly, we find that the statistical anisotropies do not depend on model parameters and become $\mathcal{O}(1)$.
Furthermore, the sourced GWs in our model are linearly polarized, in contrast to the chiral GWs discussed in the previous works.
 We expect that these fascinating signatures provide a new window to probe high energy physics through the primordial GWs and can be examined in upcoming experiments.

This paper is organized as follows. In section~\ref{Model Action and Setup}, we describe the setup of our model.
In section~\ref{Background Dynamics}, we solve the evolution of the background fields.
Then the perturbations of the spectator scalar field and the gauge field
are calculated in section~\ref{Perturbations of Spectator Fields}.
We study how they source the curvature perturbation and the primordial GWs, and their detectability is discussed in section~\ref{Generation of Inflaton Perturbation and GW}. Section~\ref{Conclusion} is devoted to the conclusion of this paper.

\section{Model Action and Setup}
\label{Model Action and Setup}

In this paper, we study a spectator scalar field coupled to a $U(1)$ gauge field in the inflationary universe and calculate perturbations.
We consider the following action:
\begin{equation}
\mathcal{L} = \frac{1}{2}(\partial_\mu \phi)^2 -U(\phi) - \frac{1}{2}(\partial_\mu \sigma)^2 -V(\sigma)
-\frac{1}{4}I^2(\sigma)F_{\mu\nu}F^{\mu\nu},
\label{model action}
\end{equation}
where $\phi$ is the inflaton, $\sigma$ is a spectator scalar field and
$F_{\mu\nu}=\partial_\mu A_\nu-\partial_\nu A_\mu$ is the field strength of a $U(1)$ gauge field $A_\mu$.
$U(\phi)$ and $V(\sigma)$ are the potentials of these scalar fields.
The spectator scalar field $\sigma$ is coupled to the kinetic term of
the gauge field through $I(\sigma)$.
We decompose these fields into the backgrounds and perturbations as
\begin{equation}
\phi(t,\bm{x})=\bar{\phi}(t)+\delta\phi(t,\bm{x}),\quad 
\sigma(t,\bm{x})=\bar{\sigma}(t)+\dsig(t,\bm{x}),\quad 
A_i(t,\bm{x})=\bar{A}_i(t)+\delta A_i(t,\bm{x}),
\end{equation}
where the radiation gauge, $\bar{A}_0(t)= \partial_i A_i(t,\bm x)=0,$ is taken.
In the following discussion, we eliminated $A_0(t,\bm{x})=\delta A_0(t,\bm{x})$ by solving the gauge constraint equation.
For simplicity, we approximate the background metric by the FLRW metric. Although the background
gauge field breaks the isotropy of the universe, its energy density
is subdominant (e.g. $\mathcal{O}(10^{-5})$ times smaller than the total energy in the example in section~\ref{Detectability}). In that case, even with the FLRW background,
we can correctly calculate the statistical anisotropy of perturbations~\cite{Bartolo:2012sd}.

In this paper, we let the inflaton model unspecified and do not solve the background evolution of the inflaton. Instead, we parameterize the cosmic expansion with a constant Hubble parameter, $H\approx const.$ 
On the other hand, $V(\sigma)$ and $I(\sigma)$ need to be fixed for concrete calculations. 
For the kinetic function $I(\sigma)$, an exponential form is theoretically well motivated by high energy physics (e.g., dilatonic coupling),
\begin{equation}
I(\sigma) = e^{\sigma/\Lambda}.
\end{equation}
Regarding the potential $V(\sigma)$, we are interested in the case where the spectator scalar field $\sigma$ slowly rolls down its potential first and then gets stabilized
by a significantly large potential curvature. Thus we consider the following $V(\sigma)$ as a simple model:
\begin{align}
V(\sigma)&= \mathcal{M}^3\frac{\sigma^2}{\sigma+\Lambda}
\ \sim\ \begin{cases}
\mathcal{M}^3\sigma\qquad \quad (\sigma\gg \Lambda)  \\
\mathcal{M}^3\sigma^2/\Lambda\ \quad (\sigma\ll \Lambda)  \\
\end{cases}.
\label{potential V}
\end{align}
In $V(\sigma)$ and $I(\sigma)$, we introduce new dimensionful parameters, $\Lambda$ and $\mathcal{M}$.
The above potential $V(\sigma)$ is just a toy model in which a linear potential for $\sigma \gg \Lambda$ and a quadratic potential for $\sigma\ll \Lambda$ are smoothly connected.\footnote{This potential is negative for $\sigma<-\Lambda$, but $\bar{\sigma}$ never goes there if it has a positive and large initial value, $\bar{\sigma}(t_{\rm in})\gg\Lambda$. Under this assumption, we are free from the strong coupling problem with a small kinetic function, $\bar{I}\ll 1$. } Note that other forms of potential are also expected to provide similar dynamics
and predictions, as far as it supports the slow-roll and stabilization of $\bar{\sigma}$.

\section{Background Dynamics}
\label{Background Dynamics}

In this section, we study the dynamics of the background fields.
The model action eq.~\eqref{model action} leads to the following background equations:
\begin{equation}
\ddot{\bar \sigma}+3H\dot{\bar \sigma}+\bar{V}'= \frac{2}{\Lambda}\bar{\rho}_E,
\qquad
\frac{\dd}{\dd t}\left( a \bar{I}^2 \dot{\bar{A}}_i\right)=\ 0,
\label{BG EoM}
\end{equation}
with the energy density of the background gauge field,
\begin{equation}
\bar{\rho}_E\equiv \frac{\bar{I}^2}{2a^2}\dot{\bar{A}}_i^2.
\end{equation}
Here, $\bar{I}\equiv I(\bar{\sigma})$ is the background kinetic function, and dot and prime denote the cosmic time derivative and the derivatives with respective to fields (e.g., $\bar{V}'\equiv \partial_\sigma V(\bar{\sigma})$), respectively. 
The equation of motion (EoM) for $\bar{A}_i$ can be integrated and one finds
$\bar{\rho}_E \propto a^{-4} \bar{I}^{-2}$.
Thus the evolution of $\bar{\rho}_E$ is simply determined by $\bar{\sigma}(t)$.

As we see below, the background evolution has the following  three phases. (i)  Growing phase:
Since its energy density is negligibly small, the gauge field contribution to the EoM of $\bar{\sigma}$  can be ignored, $|\bar{V}'|\gg 2\bar{\rho}_E/\Lambda$. The slow-roll (terminal) velocity of $\bar{\sigma}$ is solely determined by $\bar{V}'$. 
Then the kinetic energy of $\bar{\sigma}$ is transferred to the gauge field and
$\bar{\rho}_E$ increases.
(ii) Attractor phase:
As $\bar{\rho}_E$ grows, the contribution from the gauge field to the EoM of $\bar \sigma$ becomes no longer negligible. Then the velocity of $\bar{\sigma}$ slows down and the decelerated variation of kinetic function makes the energy flow to the gauge field balanced. Consequently, $\bar{\rho}_E$ stays constant. (iii) Damping phase: When $\bar{\sigma}$ reaches $\Lambda$, it starts damped oscillations due to its quadratic potential. Since $\bar{I}$ practically stops evolving,  $\bar{\rho}_E$ rapidly decays as $a^{-4}$.

Approximate solutions for these three phases can be found from the EoM as follows. In the slow-roll regime of $\sigma$ in which $\bar{\sigma}\gg\Lambda$, approximating $\bar{V}'\simeq \mathcal{M}^3$ and $\ddot{\sigma}\simeq 0$ in eq.~\eqref{BG EoM}, one finds the analytic solution of the EoM as
\begin{equation}
\bar{\sigma}(t)=\sigma_{\rm in}-\frac{\mathcal{M}^3}{3H}(t-t_{\rm in})+\frac{\Lambda}{2}\ln
\left[1+\frac{2\bar{\rho}_E(t_{\rm in})}{3\dn H^2\Lambda^2}
\left(\left(\frac{a}{a_{\rm in}}\right)^{2\dn}-1\right) \right],
\label{bs solution}
\end{equation}
where an exponentially decaying term is neglected, subscript ``in'' denotes the initial value, and we introduce an almost constant parameter ~$n$ ~defined as
\begin{equation}
n\equiv\frac{\mathcal{M}^3}{3H^2\Lambda},
\qquad
\dn\equiv n-2.
\end{equation}
Here we assume that $\bar{\rho}_E$ is set to be negligibly small at the initial time by some mechanisms.
For $\dn>0$, the term proportional to $\bar{\rho}_E(t_{\rm in})a^{2\dn}$,  which is initially negligible, eventually dominates the logarithm term in eq.~\eqref{bs solution} and it causes the shift from the growing phase into the attractor phase.
For $\bar{\sigma}\lesssim \Lambda$, however, the kinetic function stops evolving $\bar{I}\simeq 1$ and the effective mass of $\bar{\sigma}$ is given by
\begin{equation}
\bar{V}''\simeq \frac{2\mathcal{M}^3}{\Lambda} =  6n H^2 \qquad (\sigma\lesssim \Lambda).
\end{equation}
Therefore, assuming $n>2$ and $\bar{\rho}_E$ is initially small,  we find
the three phases of the background evolution,
\begin{align}
\dot{\bar \sigma}(t) &\simeq -2H \Lambda\times
\begin{cases} n/2 & (t<t_A) \\
1 & (t_A<t<t_D) \\
(a/a_D)^{-3/2}\cos(\sqrt{6n}H t +\varphi) & (t_D<t) \\
\end{cases},
\label{s0 evolution}
\\
\bar{\rho}_E(t) &\simeq \frac{3}{2}\dn H^2 \Lambda^2\times
\begin{cases}  \,(a/a_A)^{2\dn} & (t<t_A) \\
1 & (t_A<t<t_D) \\
(a/a_D)^{-4} & (t_D<t) \\
\end{cases},
\label{rhoE evolution}
\end{align}
where $t_A$ and $t_D$ are the time when $\bar \rho_E$ reaches the attractor value $\frac{3}{2}\dn H^2 \Lambda^2$ and $\bar \sigma$ reaches $\Lambda$, respectively. We denote the values of the scale factor at these times by $a_A\equiv a(t_A)$ and $a_D\equiv a(t_D)$.
$\varphi$ is a constant phase of the damped oscillation of $\bar \sigma$.

The validity of these approximate solutions for the three phases of the background dynamics can be confirmed by a numerical calculation.
The background EoMs can be recast into a dimensionless form,
\begin{align}
\partial_N^2 S+3\partial_N S+3n\,\frac{S^2+2S}{(S+1)^2}&=e^{-2N+2S}\mathcal{E}^2,
\notag\\
\partial_N \bm{\mathcal{E}}+(2\partial_N S+1)\bm{\mathcal{E}}&=0.
\end{align}
with the redefined variables $S\equiv \bar{\sigma}/\Lambda,\, \bm{\mathcal{E}}\equiv \dot{\bar{\bm A}}/H\Lambda$, and $a=e^{N}$.
From these equations, it is clear that $n$ is a unique parameter characterizing the evolution of the background system.
%
\begin{figure}[tbp]
    \hspace{-2mm}
  \includegraphics[width=70mm]{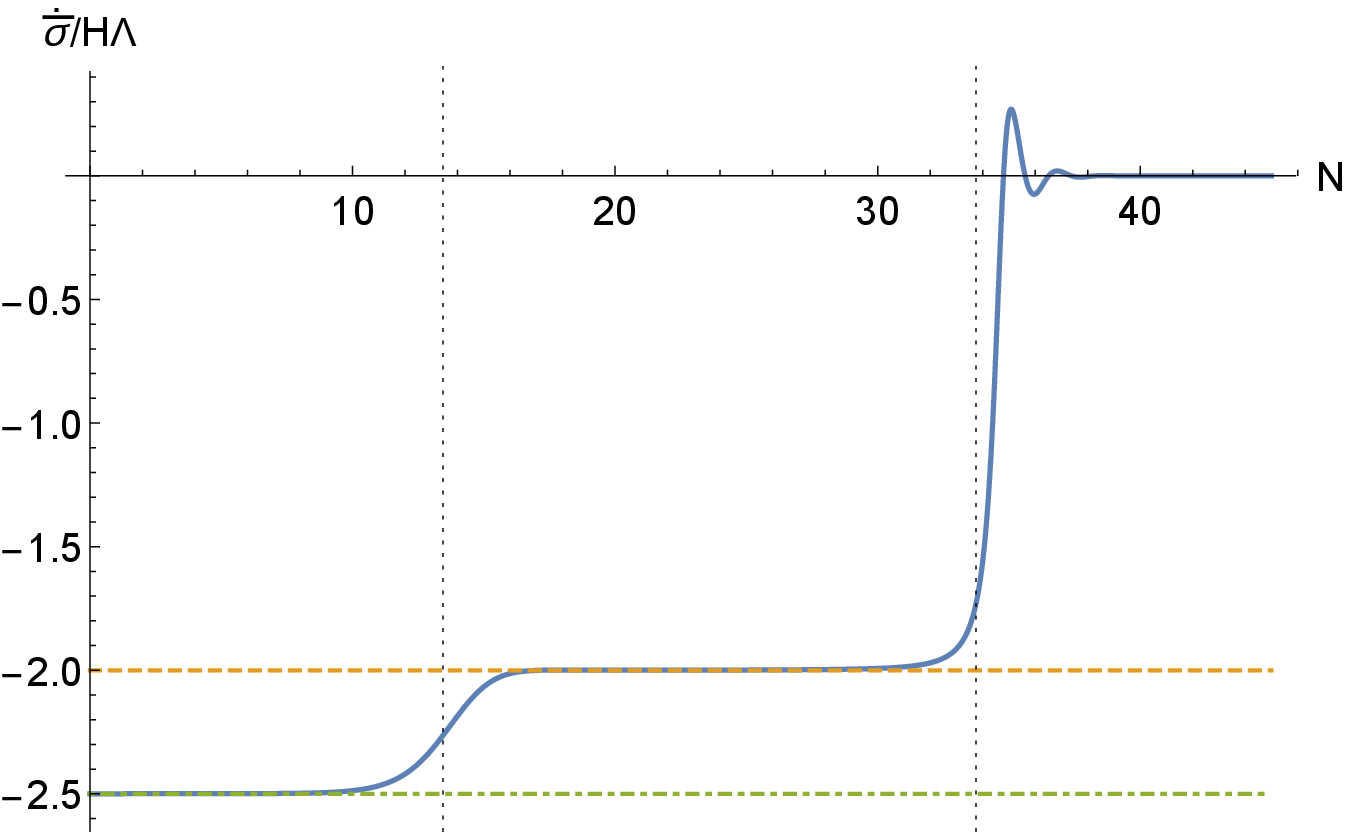}
  \hspace{5mm}
  \includegraphics[width=70mm]{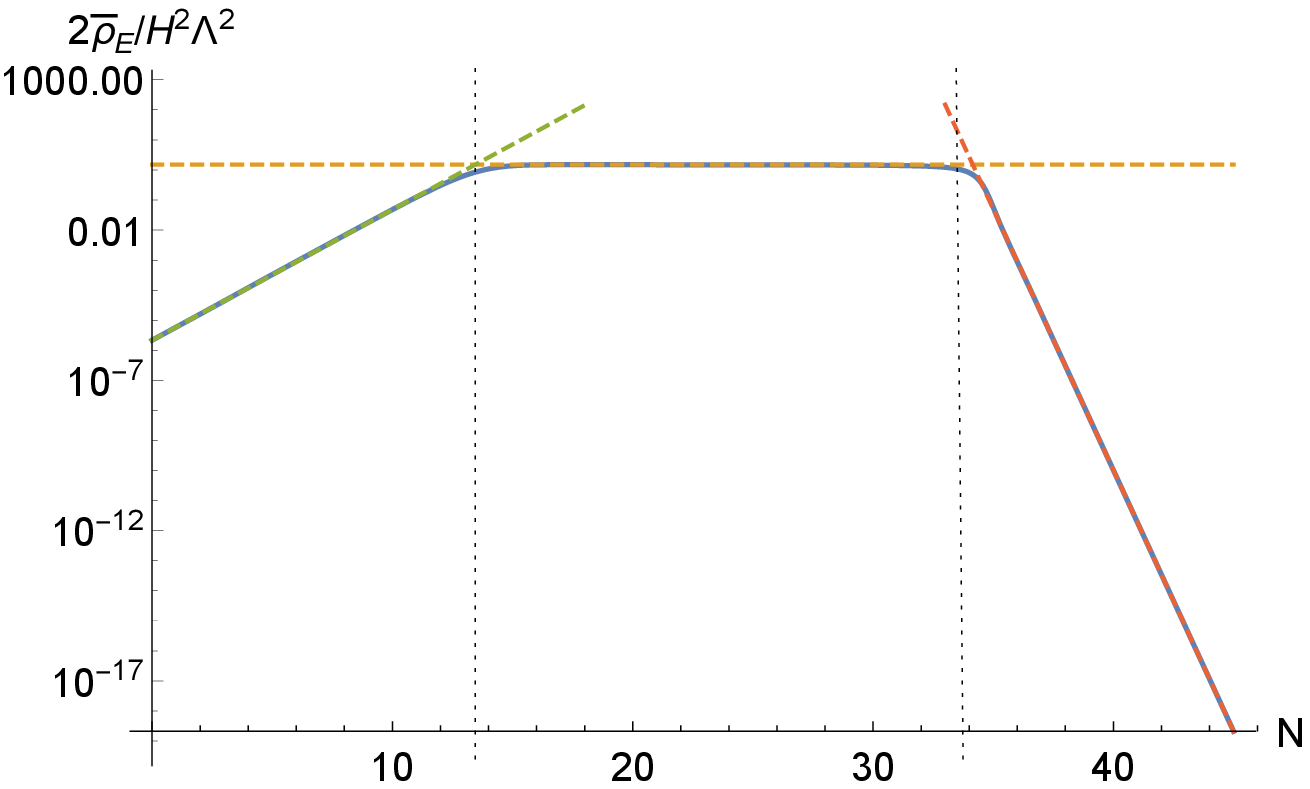}
  \caption
 {A numerical result of the time evolution of $\dot{\bar{\sigma}}$ (left panel) and $\bar{\rho}_E$ (right panel). The horizontal axis is e-folding number $N\equiv \ln (a/a_{\rm in})$. We set $n=2.5$ and the initial condition 
$\bar{\sigma}=75\Lambda, \dot{\bar{\sigma}}=-nH\Lambda, \bar{\rho}_E=10^{-6}H^2\Lambda^2$
 at $N=0$.
The green dot-dashed lines represent the analytic solutions in the growing phase, $\dot{\bar \sigma}=-nH\Lambda$ (left panel) and $10^{-6}H^2\Lambda^2\,a$ (right panel).
One can also see the analytic solutions in the attractor phase, $\dot{\bar \sigma}=-2H\Lambda$ and $\bar{\rho}_E = \frac{3}{2}\dn H^2\Lambda^2$, 
which are shown as yellow dashed lines are realized.
The transition times between the phases are illustrated as the vertical black dashed lines.
 The red dashed line in the right panel indicates $\bar{\rho}_E$ decays as $a^{-4}$ in the damping phase.}
 \label{BG numerical}
\end{figure}
%
In figure~\ref{BG numerical}, we show the numerical evaluation of $\bar{\sigma}(t)$ and $\bar{\rho}_E(t)$ with $n=2.5$, and confirm that the analytically derived behaviors are indeed realized.

\section{Perturbations of Spectator Fields}
\label{Perturbations of Spectator Fields}

In this section, we discuss $\delta \sigma$ and $\delta A_i$.
We quantize them, numerically solve their EoMs, and find approximate analytic solutions.
We mainly consider the modes which exit the horizon during the growing phase $(t<t_A)$, because the modes on smaller scales are never amplified
and it is harder for these modes to leave an observable imprint as we see in the next section.

\subsection{Quantization and numerical calculation}

\begin{figure}[tbp]
 \begin{center}
  \includegraphics[width=90mm]{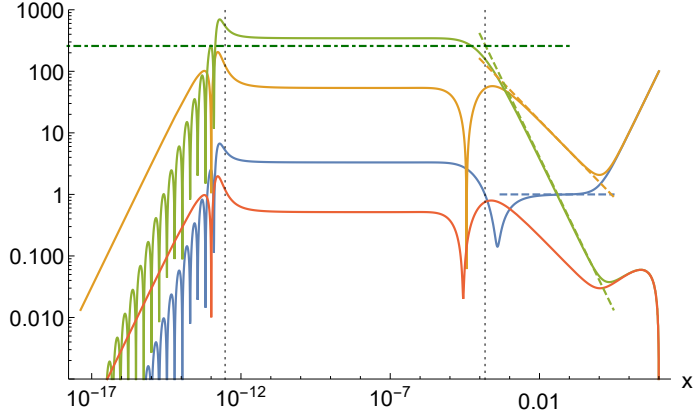}
 \end{center}
  \caption
 {The numerical results of $\sqrt{2k}x|a\delta\sigma^{\rm int}|$ (blue), $\sqrt{2k}x|\bar{I}\delta A^{\rm int}|$ (yellow), $\sqrt{2k}x|a\delta\sigma^{\rm src}|$ (green) and $\sqrt{2k}x|\bar{I}\delta A^{\rm src}|$ (red) are shown. The horizontal axis is $x\equiv -k\eta$. We fix $\theta =0.3\pi$. These modes exit the horizon before the background system enters the attractor phase at $x_A=1.4\times 10^{-4}$ and the damping phase at $x_D=3\times 10^{-13}$ (vertical black dashed lines). The other dashed lines in the figure are analytically derived in section~\ref{Analytic solutions}.}
 \label{PT numerical}
\end{figure}

We first decompose $\delta A_i$ with the linear polarization vectors $e^X_i(\hat{\bm k})$ and $e^Y_i(\hat{\bm k})$ (see appendix~\ref{Polarization Vector and Tensor} for their definition) in Fourier space as
\begin{equation}
\delta A_i(t,\bm x)= 
\int \frac{\dd^3 k}{(2\pi)^3} e^{i\bm{k}\cdot\bm{x}}\, \left[e^X_i(\hat{\bm k}) \delta A_{\bm k}^X(t)+i e^Y_i(\hat{\bm k}) \delta A_{\bm k}^Y(t)\right].
\end{equation}
Without loss of generality, we can assume that the background electric field is parallel to the $z$-axis,
\begin{equation}
\dot{\bar{\bm A}} \propto \hat{\bm z}.
\label{E prop z}
\end{equation}
In that case, the inner product between the background electric field and the polarization vector is
\begin{equation}
\sum_i\dot{\bar{A}}_i \,e_i^X(\hat{\bm k})=-\sin\theta
\, \frac{a}{\bar{I}}\sqrt{2\bar{\rho}_E},
\qquad  
\sum_i\dot{\bar{A}}_i \,e_i^Y(\hat{\bm k})=0,
\label{inner product}
\end{equation}
where $\cos\theta\equiv \bm{k}\cdot \dot{\bar{\bm A}}/(|\bm k| |\dot{\bar{\bm A}}|)$.
Only the $X$ mode can make a scalar, combined with the background electric field, and hence is coupled to $\delta\sigma_{\bm k}$ 
in the quadratic action.
We first focus on the $X$ mode and neglect $\delta A^Y_k$ for a while.
The Fourier transformations of $\delta\sigma$ is as usual,
\begin{equation}
\delta \sigma(t,\bm x)= 
\int \frac{\dd^3 k}{(2\pi)^3} e^{i\bm{k}\cdot\bm{x}}
\delta \sigma_{\bm k}(t).
\end{equation}
The quadratic action of $\delta\sigma_{\bm k}(\eta)$ and $\delta A_{\bm k}^X(\eta)$ without the gravitational coupling and slow-roll corrections is given by
(the full expression can be found in appendix~\ref{Quadratic Action})
\begin{equation}
S^{(2)}_\Delta=\frac{1}{2}\int\dd\eta\frac{\dd^3 k}{(2\pi)^3}\left[
\partial_\eta\Delta^\dag \partial_\eta\Delta+ \partial_\eta\Delta^\dag K \Delta- \Delta^\dag K \partial_\eta\Delta
-\Delta^\dag \Omega^2 \Delta\right],
\end{equation}
with 
\begin{align}
&\Delta
=\begin{pmatrix}a\dsig_{\bm k} \\
\bar{I} \delta A_{\bm k}^X \\
\end{pmatrix},
\quad
K=\frac{\sqrt{2\bar{\rho}_E}}{\Lambda H\eta}\sin\theta\begin{pmatrix}0 & -1 \\
1 & 0 \\
\end{pmatrix},
\notag\\
&\Omega^2 = \begin{pmatrix}k^2-(2-\mu_\sigma^2/H^2)/\eta^2 & 
\sqrt{2\bar{\rho}_E}\sin\theta \partial_\eta(\ln[\bar{I}/a])/(\Lambda H \eta) \\
\sqrt{2\bar{\rho}_E}\sin\theta \partial_\eta(\ln[\bar{I}/a])/(\Lambda H \eta)
& k^2-\partial_\eta^2 \bar{I}/\bar{I}\\
\end{pmatrix},
\end{align}
where $\eta$ is the conformal time and $\mu_\sigma^2= \bar{V}''+ 4\Lambda^{-2}\bar{\rho}_E\cos(2\theta)$.
With these expressions, the EoMs are given by
\begin{equation}
\partial_\eta^2 \Delta+2K \partial_\eta \Delta+(\Omega^2 + \partial_\eta K)\Delta=0,
\label{Matrix EoM}
\end{equation}
%
Since this system has both kinetic mixing and mass mixing, the coupled EoMs cannot be diagonalized. Hence we solve the evolution of four modes which are the perturbations of $\dsig$ and $\delta A$ originating from the vacuum fluctuation of the respective fields. 
Promoting $\Delta$ into operators as~\cite{Dimastrogiovanni:2012ew}
\begin{equation}
\hat{\Delta}
=\begin{pmatrix} a\delta\sigma^{\rm int}_k & a\delta\sigma^{\rm src}_k \\
\bar{I}\delta A_k^{\rm src} & \bar{I}\delta A_k^{\rm int} \\
\end{pmatrix}
\begin{pmatrix} \hat{a}_{\bm k} \\
\hat{b}_{\bm k} \\
\end{pmatrix}+{\rm h.c.}.
\label{Matrix quantization}
\end{equation}
%
The quantization is done by imposing the standard commutation relations
to two independent sets of annihilation/creation operators, $\{\hat{a}_{\bm k}, \hat{a}^\dag_{\bm k}\}$ and $\{\hat{b}_{\bm k}, \hat{b}^\dag_{\bm k}\}$.
The subscripts ``int'' and ``src'' represent the intrinsic modes and the sourced modes, respectively. 
Since $a\delta\sigma$ and $\bar{I}\delta A$ are decoupled in the sub-horizon limit, it is reasonable to assume that $a\delta\sigma_k^{\rm int}$ and $\bar{I}\delta A_k^{\rm int}$ are identical to the one for the  Bunch-Davies vacuum in the distant past, while $a\delta\sigma_k^{\rm src}$ and $\bar{I}\delta A_k^{\rm src}$ vanish there:
\begin{equation}
\lim_{|k\eta|\to \infty}
\begin{pmatrix}a\delta\sigma_k^{\rm int}(\eta) & a\delta\sigma_k^{\rm src}(\eta) \\
\bar{I}\delta A_k^{\rm src}(\eta) & \bar{I}\delta A_k^{\rm int}(\eta) \\
\end{pmatrix}=\frac{e^{-ik\eta}}{\sqrt{2k}}\begin{pmatrix}1 & 0 \\
0 & 1 \\
\end{pmatrix}.
\end{equation}
Finally we obtain the EoMs for the mode functions as
\begin{multline}
\partial_x^2\begin{pmatrix} a\delta\sigma_k^{\rm int} & a\delta\sigma_k^{\rm src} \\
\bar{I}\delta A_k^{\rm src} & \bar{I}\delta A_k^{\rm int} \\
\end{pmatrix}
+\frac{2\sqrt{2\bar{\rho}_E}}{\Lambda Hx}\sin\theta
\begin{pmatrix}0 & -1 \\
1 & 0 \\ \end{pmatrix}
\partial_x\begin{pmatrix} a\delta\sigma_k^{\rm int} & a\delta\sigma_k^{\rm src} \\
\bar{I}\delta A_k^{\rm src} & \bar{I}\delta A_k^{\rm int} \\
\end{pmatrix}
\\+
\begin{pmatrix}1-\frac{2-\mu_\sigma^2/H^2}{x^2} & 
\frac{2\sqrt{2\bar{\rho}_E}\sin\theta}{\Lambda^2 H x} \partial_x \bar{\sigma} \\
\frac{2\sqrt{2\bar{\rho}_E}\sin\theta}{\Lambda H x^2}
\quad
& 1- \partial_x^2 \bar{\sigma}/\Lambda
-\left(\partial_x \bar{\sigma}/\Lambda\right)^2\\
\end{pmatrix}
\begin{pmatrix} a\delta\sigma_k^{\rm int} & a\delta\sigma_k^{\rm src} \\
\bar{I}\delta A_k^{\rm src} & \bar{I}\delta A_k^{\rm int} \\
\end{pmatrix}=0,
\label{Component EoM}
\end{multline}
where $x\equiv -k\eta$ is introduced as a new time variable.
The $x$ derivatives of the background scalar field $\bar{\sigma}(t)$ can be rewritten as
\begin{equation}
\partial_x\bar{\sigma}=-\frac{\dot{\bar \sigma}}{H x},
\qquad
\partial_x^2\bar{\sigma}=\frac{\ddot{\bar \sigma}+H\dot{\bar \sigma}}{H^2 x^2} \simeq \frac{\dot{\bar \sigma}}{H x^2}.
\end{equation}
It should be noted that all the off-diagonal terms are proportional to $\sin\theta$
with $\theta$  being the angle between $\bm k$ and the background gauge field $\dot{\bar{\bm A}}$ (see eq.~\eqref{inner product}). This can be understood as follows. Scalar and vector perturbations
are decoupled at first order in isotropic background by virtue of the decomposition theorem, but the background vector field which (weakly) breaks the isotropy of the universe enables their coupling in our model. When $\bm k$ is parallel to $\dot{\bar{\bm A}}$, this coupling disappears.
We numerically solve the above coupled equations
for modes that exit the horizon during the growing phase.
In figure~\ref{PT numerical}, we show the numerical results.
%
%
In the next subsection, we develop an analytic treatment to understand the numerical results.

\subsection{Analytic solutions}
\label{Analytic solutions}

The EoMs of $\delta\sigma_k$ and $\delta A^X_k$, respectively, are given by
\begin{align}
\left[\partial_x^2+1-\frac{2-\mu_\sigma^2/H^2}{x^2}\right](a\delta\sigma_k)
&=2\sqrt{2}\sin\theta \frac{\sqrt{\bar{\rho}_E}}{H\Lambda\, x}\,\bar{I}\partial_x \delta A_k^X
,
\\
\left[\partial_x^2 +1-\frac{\partial_x^2 \bar{I}}{\bar{I}}\right](\bar{I}\delta A_k^X)  &=
-2\sqrt{2} \sin\theta \frac{\sqrt{\bar{\rho}_E}}{H\Lambda\,x}\, a\partial_x \delta\sigma_k,
\label{dA EoM}
\end{align}
where $\bar{V}''$ in $\mu_\sigma^2 \equiv \bar{V}''+ 4\Lambda^{-2}\bar{\rho}_E\cos(2\theta)$ can be ignored during the growing and attractor phases.
Then, although $\mu_\sigma^2$ is negative for $\pi/4<\theta<3\pi/4$,
it does not lead to tachyonic instability  as we see soon.

\subsubsection{Growing phase}

During the growing phase, since $\bar{\rho}_E\ll H^2\Lambda^2$, all the terms with $\bar{\rho}_E$ including the coupling terms between $\delta\sigma$ and $\delta A$ are sub-leading.
Then it is straightforward to obtain the homogeneous solutions in the super-horizon limit as,
\begin{equation}
a\delta\sigma^{\rm int}_k
\simeq
\frac{i }{\sqrt{2k}\,x}, \qquad \bar{I} \delta A^{\rm int}_k \simeq \frac{\Gamma(n-\frac{1}{2})}{\sqrt{2\pi k}}\left(\frac{x}{2}\right)^{1-n},
\qquad  (x\ll 1).
\label{ds before attractor}
\end{equation}
They are plotted as the blue and yellow dashed lines in figure~\ref{PT numerical}. Note that $\bar{I} \delta A^{\rm int}_k/a$ becomes much larger than $\delta \sigma^{\rm int}_k$, because the former grows on super-horizon scales in proportion to $a^{n-2}$, while the latter stays constant.
We do not discuss $\bar{I}\delta A^{\rm src}_k$, which is sourced by $a\delta\sigma^{\rm int}_k$ and hence sub-leading (see the red line in figure~\ref{PT numerical}).

$a \delta\sigma^{\rm src}_k$ sourced by  $\bar{I}\delta A_k^{\rm int}$ on super-horizon scales during the growing phase can be obtained with the Green's function method. $\delta\sigma^{\rm src}_k$ can be calculated as
\begin{equation}
a\delta\sigma^{\rm src}_k(x)
=2\sqrt{2}\sin\theta\int\dd y\, G_{R}(x,y) \,  \frac{\sqrt{\bar{\rho}_E(y)}}{H\Lambda\, y}\,\bar{I}\partial_y \delta A^{\rm int}_k(y),
\label{G function}
\end{equation}
The retarded Green's function $G_{R}(x,y)=-\Theta(y-x)\,(x^3-y^3)/(3xy)$ satisfies $\left[\partial_x^2-2/x^2\right]G_R (x,y)=\delta(x-y)$
in which the gradient term and the  mass term $\mu_\sigma^2$ are ignored. Substituting 
$\rho_E(x)\simeq\frac{3}{2}\Delta n\, H^2\Lambda^2 (x/x_A)^{-2\dn}
$ from eq.~\eqref{rhoE evolution}\footnote{
We define $x_A$ as the time when $\bar{\rho}_E(t_{\rm in})\left(a/a_{\rm in}\right)^{2(n-2)}$ reaches the attractor value $\frac{3}{2}\dn H^2 \Lambda^2$. Then one finds $x_A=x_{\rm in}[\bar{\rho}_E(t_{\rm in})/(\frac{3}{2}\dn H^2 \Lambda^2)]^{\frac{1}{2\dn}}$.}
and integrating eq.~\eqref{G function}, we obtain
\begin{equation}
a\delta\sigma^{\rm src}_k\simeq \frac{1}{\sqrt{2k}x}\, \sqrt{\frac{3}{\pi \dn}}
\frac{2^n\Gamma(n+1/2)}{2n-1}\sin\theta\, \frac{x^{4-2n}}{x_A^{2-n}}.
\qquad  (x\ll 1)
\label{sigma src}
\end{equation}
It is plotted in figure~\ref{PT numerical} as a green dashed line.
Therefore $\delta \sigma^{\rm src}$ grows as $a^{2n-4}$ on super-horizon scales during the growing phase, faster than $\bar{I}\delta A^{\rm int}/a$.

\subsubsection{Attractor phase}

\begin{figure}[tbp]
    \hspace{-2mm}
  \includegraphics[width=70mm]{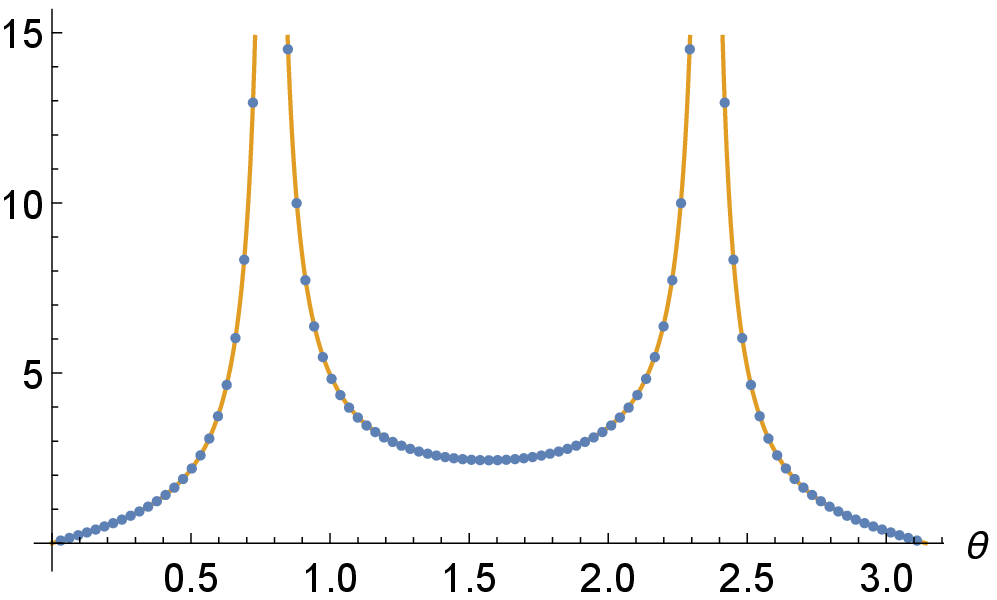}
  \hspace{5mm}
  \includegraphics[width=70mm]{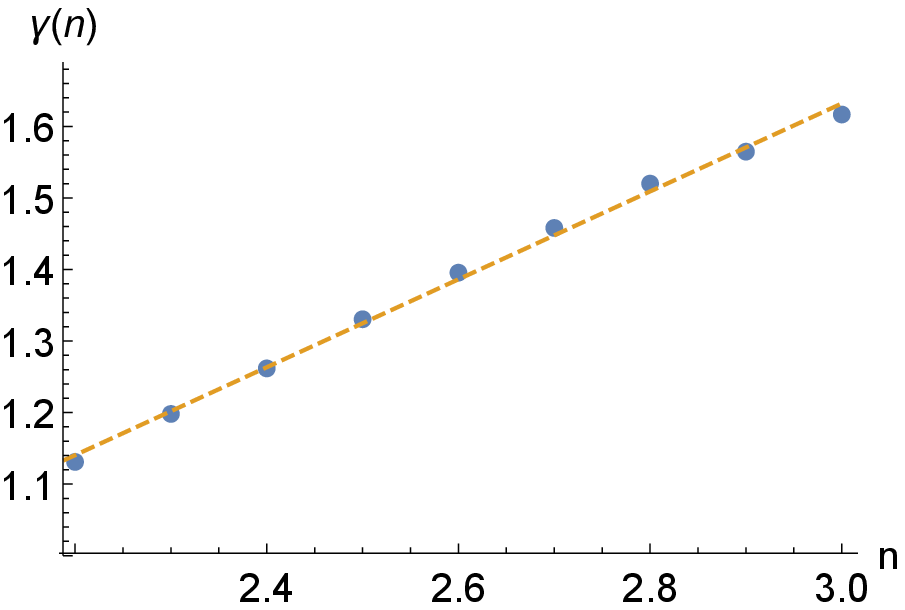}
  \caption
 {{\bf (Left panel)} $a\delta\sigma_k/\bar{I} \delta A_k^X$ analytically derived in eq.~\eqref{super horizon ratio} (yellow line) and numerically obtained $a\delta\sigma^{\rm src}_k/\bar{I} \delta A^{\rm int}_k$ during the attractor phase (blue dots) are compared. The setting of the numerical calculation is the same as figure~\ref{PT numerical}. An excellent agreement is seen. {\bf (Right panel)} $\gamma(n) $ is the numerically computed $\delta \sigma^{\rm src}$ in the super-horizon limit during the attractor phase divided by the analytic expression eq.~\eqref{analytic dsig}. The blue dots are the numerical result and the yellow line is a linear fitting function, eq.~\eqref{linear fit}.}
 \label{ratio and gamma}
\end{figure}

We can derive a simple relationship between $a\delta\sigma_k$ and $\bar{I}\delta A_k^X$ on super-horizon scales during the attractor phase.
Changing the time variable from conformal time to cosmic time, one can rewrite
the EoM of $\delta\sigma$ as 
\begin{align}
\ddot{\delta\sigma}_k+3H\dot{\delta\sigma}_k+6\Delta n\cos(2\theta)H^2\delta\sigma_k
=-2
\sqrt{3\Delta n}  \sin\theta\ H\frac{\bar{I}\delta\dot{A}_k^X}{a},
\label{att sig EoM}
\end{align}
where the spatial gradient terms are ignored and some background time dependence during the attractor phase is used (see appendix~\ref{Super-horizon solutions during attractor phase} for derivation).
As shown in appendix~\ref{Super-horizon solutions during attractor phase},
we find that $\delta\sigma_k$ and $\bar{I}\delta A_k^X/a$ have a constant solution while the others are decaying.
Focusing on the constant solution ($\delta\sigma=\text{const}, \bar{I}\delta A_k^X\propto a$), we find the following simple relation
which depends only on $n$ and $\theta$:
\begin{equation}
\frac{a\delta\sigma_k}{\bar{I} \delta A_k^X}=
-\sqrt{\frac{3}{\Delta n}}\frac{\sin\theta}{\cos 2\theta},
\qquad {\rm (super\ horizon)}
\label{super horizon ratio}
\end{equation}
The EoM for $\bar{I}\delta A_k^X$ is trivially satisfied in this limit.
Note that the pairs of modes coupled through eq.~\eqref{att sig EoM}, i.e., $\{a\delta\sigma^{\rm src}_k,\, \bar{I} \delta A^{\rm int}_k\}$ and $\{a\delta\sigma^{\rm int}_k,\,\bar{I} \delta A^{\rm src}_k\}$, satisfy this relation.
In the left panel of figure~\ref{ratio and gamma}, we numerically confirm this relationship.
%
%

Now one needs to connect the solutions during the attractor phase
to the one during the growing phase to determine the amplitude
of $a\delta\sigma_k$ or $\bar{I}\delta A_k^X$. 
It is difficult to obtain the exact analytic solution of $a\delta\sigma_k$ and $\bar{I}\delta A_k^X$ around the transition between the growing and the attractor phase for the following two reasons: (i) The background dynamics is not simple. For instance,
$\bar{\rho}_E$ changes its time evolution from $\bar{\rho}_E\propto a^{2\dn}$ to $\bar{\rho}_E=$ const. (ii) The coupling between the perturbations gradually increases.

Thus we use a rather crude approximation. We extrapolate $\delta\sigma^{\rm src}_k$ of the growing phase till the transition time $x=x_A\equiv -k\eta_A$. Substituting $x=x_A$ into eq.~\eqref{sigma src}, we obtain
\begin{equation}
\delta\sigma^{\rm src}_k\sim \frac{H}{\sqrt{2k}k}\, \sqrt{\frac{3}{\pi \dn}}
\frac{2^n\Gamma(n+1/2)}{2n-1}\sin\theta\, \left(\frac{k_A}{k}\right)^{\dn}.
\label{analytic dsig}
\end{equation}
where we rewrite $x_A^{2-n}=\left(k_A/k\right)^{\dn}$ and $k_A$ is the wave number which exits horizon when the background enters the attractor phase.
This expression is plotted in figure~\ref{PT numerical} as a dark green dot-dashed line.
As expected, one can see a small deviation from the numerical result.
To compensate this $\mathcal{O}(1)$ discrepancy, we introduce
a factor $\gamma(n)$ which is numerically computed and obtain the constant amplitude of $\delta\sigma^{\rm src}$ on super-horizon scale as
\begin{equation}
\delta\sigma^{\rm src}_k= \frac{H}{\sqrt{2k}k}\,\tilde{\gamma}(n)\sin\theta\left(\frac{k_A}{k}\right)^{\dn},
\qquad
\tilde{\gamma}(n)\equiv \gamma(n)\sqrt{\frac{3}{\pi \dn}}
\frac{2^n\Gamma(n+1/2)}{2n-1}.
\end{equation}
Using the relation~\eqref{super horizon ratio},
we also obtain the constant amplitude of $\bar{I}\delta A^{\rm int}/a$ as well,
\begin{equation}
a^{-1}\bar{I}\delta A^{\rm int}_k= -\frac{H}{\sqrt{2k}k}\,\sqrt{\frac{\dn}{3}}\tilde{\gamma}(n)\cos2\theta\left(\frac{k_A}{k}\right)^{\dn}.
\label{AX solution}
\end{equation}
Both $\delta\sigma^{\rm src}_k$ and $\delta A^{\rm int}_k$ have red-tilted spectrum,
because they continue to grow from the horizon exit until the attractor phase starts.

We numerically calculate $\gamma(n)$ and show it in the right panel of figure~\ref{ratio and gamma}.
For $2.2\le n\le 3$, we found a linear fitting function,\footnote{As $n$ is closer to 2, the transition from the growing phase into the attractor phase takes longer time. Then the damping phase may start or even the observable inflation may end before the system enters the attractor phase. The former case was studied in ref.~\cite{Choi:2015wva}, while we assume the attractor phase exists in this paper.}
\begin{equation}
\gamma(n)\approx 0.614n-0.211.
\qquad (2.2\le n\le 3)
\label{linear fit} 
\end{equation}
Considering that eq.~\eqref{analytic dsig} is highly sensitive to $n$ (e.g., eq.~\eqref{analytic dsig} with $n=2.5$ and $n=3$ are different by a factor of more than $100$ for $x_A\approx 1.4\times 10^{-4}$), the weak dependence of $\gamma(n)$ on $n$ implies that eq.~\eqref{analytic dsig} is a reasonable approximation.
It is also numerically checked that $\gamma(n)$ does not depend on $\theta$.

\subsubsection{Damping phase}

Since a perturbation on super-horizon scales behaves in the same way as its background component, $\delta \sigma_k$  oscillates with an amplitude decaying as $a^{-3/2}$ and the electric component $\bar{I} \delta \dot{A}_k/a$ decays as $a^{-2}$, which are indeed confirmed in figure~\ref{PT numerical}.
Small bumps of $\delta \sigma_k$ and $\bar{I}\delta A_k/a$ are also seen when the background enters the damping phase.
However, if the attractor phase lasts for a sufficiently long time,
the contributions from the small bumps to the sourced inflaton perturbation and GWs are negligible. 
In the following sections, we calculate the generation of $\delta\phi$ and $h_{ij}$
by focusing on the attractor phase.

\section{Generation of Inflaton Perturbation and GWs}
\label{Generation of Inflaton Perturbation and GW}

In this section, we study the generation of perturbations, the inflaton 
$\delta\phi,$ and GWs $h_{ij}$, sourced by $\delta\sigma$ and $\delta A_i$ during the attractor phase. Since the background fields and their perturbations in the spectator sector quickly decay during inflation in our scenario, they are not observable directly. Nonetheless,
the curvature  and the GW perturbations sourced by them may be observed.
We mainly discuss the modes which exit the horizon during the growing phase, because these perturbations of the spectator fields are amplified on super-horizon scales during the growing phase, and  give a sizable sourcing effect. 

\subsection{Sourced inflaton perturbation}

The EoM for $\delta\phi(t,\bm x)$ is given by
\begin{equation}
\left[\partial_t^2+3H\partial_t-\frac{\nabla^2}{a^2}+\mu_\phi^2\right]\delta\phi
=-\Omega_{\phi\sigma}\delta\sigma-\Omega_i^{A\phi}\frac{\bar{I}^2}{a^2}\delta A_i,
\label{dphi EoM}
\end{equation}
where the full expressions for $\mu_\phi^2,\ \Omega_{\phi\sigma}$ and $\Omega_i^{A\phi}$
can be found in appendix~\ref{Quadratic Action}.
Since we are interested in a super-horizon mode sourced by ~$\delta\sigma$ ~and ~$\delta A_i$ ~during attractor phase, eq.~\eqref{dphi EoM} can be reduced into
\begin{equation}
\left[\partial_x^2-\frac{2}{x^2}\right](a\delta\phi_k)
=-\frac{\Omega_{\phi\sigma}}{x^2H^2}\,a\delta\sigma^{\rm src}_k,
\end{equation}
where we have ignored the gradient term, the inflaton mass and the contribution
from the gauge field, because $\Omega^{A\phi}_i$ is suppressed by slow-roll parameters compared to $\Omega_{\phi\sigma}$ 
and $a\delta\sigma^{\rm src}$ and $\bar{I}\delta A_k^{\rm int}$
are the same order due to the relation eq.~\eqref{super horizon ratio}.
We also used $\delta \sigma_k^{\rm src}\gg \delta \sigma_k^{\rm int}$, since we are interested in the perturbations on scales where $\delta\sigma^{\rm src}_k$ is amplified significantly during the growing phase (see figure~\ref{PT numerical}).
During the attractor phase, the coupling between $\delta\phi$ and $\delta\sigma$ is rewritten as
\begin{equation}
\Omega_{\phi\sigma} \simeq
-\frac{\dot{\bar \phi}\dot{\bar \sigma}}{\Mpl^2}\left[3-\frac{\bar{I} \bar{I}' \dot{\bar{A}}_i^2}{a^2 H\dot{\bar \sigma}}\sin^2\theta\right] = 3n\sqrt{2\epsilon_\phi}\,H^2\frac{\Lambda}{\Mpl}\left(1-\frac{\dn}{n}\cos^2\theta\right),
\end{equation}
where $\dot{\bar \phi}/\Mpl H\simeq \sqrt{2\epsilon_\phi}$ is used.
Assuming $\epsilon_\phi\approx const.$, we obtain the sourced inflaton perturbation as
\begin{align}
a\delta \phi^{(s)}= -\frac{\Omega_{\phi\sigma}}{H^2}\,\delta\sigma^{\rm src}\int \dd y \, G_R(x,y) \frac{a}{y^2}
= -\frac{\Omega_{\phi\sigma}}{3H^2}\,a\delta\sigma^{\rm src}(N_A-1/3),
\end{align}
where we have performed the time integration over only the attractor phase and $N_A$ denotes the e-fold number of the duration of the attractor phase.
Putting all together and dropping an overall minus sign, we find
\begin{equation}
\frac{\delta\phi^{(s)}}{\delta\phi^{\rm (vac)}}
=n\tilde{\gamma}(n)\sin\theta\left(1-\frac{\dn}{n}\cos^2\theta\right) \sqrt{2\epsilon_\phi} \frac{\Lambda}{\Mpl}
\left(\frac{k_A}{k}\right)^{\dn}(N_A-1/3),
\end{equation}
where the amplitude of the vacuum contribution is $\delta\phi^{\rm (vac)}=H/\sqrt{2k^3}$.
Thus, as anticipated, the sourced $\delta\phi$ is suppressed by the slow-roll parameter $\epsilon_\phi^{1/2}$
and $\Lambda/\Mpl$, while it is boosted by $\left(k_A/k\right)^{\dn}$ and $N_A$ compared to the conventional vacuum fluctuation.
The power spectrum of the sourced curvature perturbation for $k\ll k_A$ is
\begin{align}
\mcP_\zeta^{(s)}
&= \mcP_\zeta^{\rm (vac)}
\left[n\tilde{\gamma}(n)\, \sqrt{2\epsilon_\phi} \frac{\Lambda}{\Mpl}
\left(\frac{k_A}{k}\right)^{\dn}(N_A-1/3)\right]^2
\notag\\
&\quad\times
\left(1-\frac{3n-4}{n}\cos^2\theta+\frac{\dn(3n-2)}{n^2}\cos^4\theta-\frac{\dn^2}{n^2}\cos^6\theta\right),
\label{sourced Pz}
\end{align}
where $\mcP_\zeta^{\rm (vac)}\equiv H^2/(8\pi^2\Mpl^2 \epsilon_\phi)$,
which is the power spectrum of the curvature perturbation contributed only by the vacuum fluctuation of $\delta\phi$
as $\zeta_{\bm k}=-\delta\phi_{\bm k}/(\sqrt{2\epsilon_\phi}\Mpl)$.

\subsection{Sourced GWs}
\label{Sourced GW}

The EoM for GW perturbations is given by
\begin{equation}
\left[\partial_t^2+3H\partial_t-\frac{\nabla^2}{a^2}\right]h_{ij}(t,\bm x)
=-\frac{4\bar{I}^2}{a^2\Mpl^2}\left(\dot{\bar{A}}_i\delta\dot{A}_j(t,\bm x)
+\dot{\bar{A}}_i\dot{\bar{A}}_j\frac{\delta\sigma(t,\bm x)}{\Lambda}\right).
\label{h EoM}
\end{equation}
We decompose GW perturbations with the polarization tensors (see appendix~\ref{Polarization Vector and Tensor} for their definitions),
\begin{equation}
h_{ij}(t,\bm x)= 
\int \frac{\dd^3 k}{(2\pi)^3} e^{i\bm{k}\cdot\bm{x}}\, \left[e^+_{ij}(\hat{\bm k}) h_{\bm k}^+(t)+i e^\times_{ij}(\hat{\bm k}) h_{\bm k}^\times (t)\right].
\end{equation}
Then one obtains the EoMs for the sourced GW mode functions, $h^+_k(t)$ and $h^\times_k(t)$, as
\begin{align}
\left[\partial_t^2+3H\partial_t+\frac{k^2}{a^2}\right]h^+_k &=
\frac{4\sqrt{\bar{\rho}_E}}{a\Mpl^2}\sin\theta\left[ \bar{I} \delta\dot{A}^X_k
-a\sqrt{2\bar{\rho}_E}\sin\theta \frac{\dsig_k}{\Lambda}\right],
\label{h+ EoM}
\\
\left[\partial_t^2+3H\partial_t+\frac{k^2}{a^2}\right]h^\times_k &=
\frac{4\sqrt{\bar{\rho}_E}}{a\Mpl^2}\sin\theta \, \bar{I} \delta\dot{A}^Y_k,
\end{align}
where we have used the background equations during the attractor phase.
It is interesting to note that $h^+_k$ is sourced by
$\delta \dot{A}^X_k$ and $\delta \sigma_k$, while $h^\times_k$ is sourced only by $\delta \dot{A}^Y_k$.
Introducing the canonical field,
\begin{equation}
\psi_{k}^\lambda \equiv \frac{1}{2}a\Mpl h_{k}^\lambda,
\qquad (\lambda=+,\times)
\end{equation}
and changing the time variable from the cosmic time to $x\equiv -k\eta$, one rewrites eq.~\eqref{h+ EoM} in the super-horizon limit as
\begin{align}
\left[ \partial_x^2-\frac{2}{x^2}\right]\psi^+
=-\frac{3\sqrt{2}\dn}{x^2}\frac{\Lambda}{\Mpl}
\cos^2\theta \, a \delta\sigma^{\rm src},
\end{align}
where we used eqs.~\eqref{rhoE evolution} and \eqref{att sig EoM}. 
With the Green's function method, we obtain
\begin{equation}
\psi^{+}_{(s)}=-\frac{aH}{\sqrt{2k}k} \sqrt{2}\dn\tilde{\gamma}(n)\,
\sin\theta \cos^2\theta
\frac{\Lambda}{\Mpl}\left(\frac{k_A}{k}\right)^{\dn}(N_A-1/3).
\end{equation}
Thus, dropping the overall minus sign, we find that the sourced GW perturbation divided by its vacuum fluctuation is given by
\begin{equation}
\frac{\psi^{+}_{(s)}}{\psi_{\rm (vac)}}=
\sqrt{2}\dn\,\tilde{\gamma}(n)\,
\sin\theta \cos^2\theta
\frac{\Lambda}{\Mpl}\left(\frac{k_A}{k}\right)^{\dn}(N_A-1/3),
\end{equation}
where $\psi_{\rm (vac)}=aH/\sqrt{2k^3}$.

Here is an easy way to find $\psi^\times_{(s)}$ sourced by $\delta A^Y_k$.
The EoM of $\delta A^Y_k$ can be reproduced by taking the limit
$\theta \to 0$ in the EoM of $\delta A^X_k$ eq.~\eqref{dA EoM},
because the coupling to $\dsig$ vanishes in this limit.
Thus the super-horizon solution of $\delta A^Y_k$ during the attractor phase can be obtained by taking the limit $\theta\to 0$ in the solution of $\delta A^X_k$ eq.~\eqref{AX solution},
\begin{equation}
a^{-1}\bar{I}\delta A^Y_k= -\frac{H}{\sqrt{2k}k}\,\sqrt{\frac{\dn}{3}}\tilde{\gamma}(n)\left(\frac{k_A}{k}\right)^{\dn}.
\end{equation}
With this solution, we find the sourced GW perturbation is given by
\begin{equation}
\frac{\psi^{\times}_{(s)}}{\psi_{\rm (vac)}}=
\sqrt{2}\dn\,\tilde{\gamma}(n)\,
\sin\theta\,
\frac{\Lambda}{\Mpl}\left(\frac{k_A}{k}\right)^{\dn}(N_A-1/3).
\end{equation}
The power spectrum of the sourced GW perturbation for $k\ll k_A$ is 
\begin{align}
\mcP_h^{(s)}&=\frac{1}{2}\mcP_h^{\rm (vac)}\left(\left|\frac{\psi^{+}_{(s)}}{\psi_{\rm (vac)}}\right|^2+\left|\frac{\psi^{\times}_{(s)}}{\psi_{\rm (vac)}}\right|^2\right),
\notag\\
&=\frac{2H^2}{\pi^2\Mpl^2}\left(1-\cos^2\theta+\cos^4\theta-\cos^6\theta\right)
\left[\dn\,\tilde{\gamma}(n)\frac{\Lambda}{\Mpl}\left(\frac{k_A}{k}\right)^{\dn}\left(N_A-\frac{1}{3}\right)\right]^2
,
\label{sourced Ph}
\end{align}
where $\mcP_h^{\rm (vac)}=2H^2/(\pi^2 \Mpl^2)$.
It is interesting to note that the statistical anisotropy of $\mcP_h^{(s)}$, namely $(1-\cos^2\theta+\cos^4\theta-\cos^6\theta)$, does not depend on any model parameters and thus it is a unique and robust prediction of our model.
It should be also noted that the GW power spectra of the two linear polarizations have different angular dependences,
\begin{equation}
\mcP_h^{+} \propto \cos^4\theta(1-\cos^2\theta),
\qquad 
\mcP_h^{\times} \propto 1-\cos^2\theta,
\label{different SA}
\end{equation}
which is another fascinating observational signature of our model.

\subsection{Scalar-tensor cross correlation}

This model also has a non-vanishing cross-correlation between the sourced GW and  curvature perturbations.
That is calculated as
\begin{align}
\mcP_{h\zeta}(k)
&\simeq\mcP_{h^+\zeta}(k)\notag\\
&=\frac{n\dn}{\sqrt{2}\pi^2}\frac{H^2}{\Mpl^2} \left(\cos^2\theta-2\frac{n-1}{n}\cos^4\theta+\frac{\dn}{n}\cos^6\theta\right)
\left[\tilde{\gamma}(n)\frac{\Lambda}{\Mpl}\left(\frac{k_A}{k}\right)^{\dn}\left(N_A-\frac{1}{3}\right)\right]^2.
\end{align}
%
The cross-mode $h^\times_k$ sourced by $\delta A^Y_k$ is not correlated to $\zeta$ sourced by $\delta \sigma_k$ at the leading order.
Comparing it with the sourced GW power spectrum, one finds
\begin{equation}
\frac{\mcP_{h\zeta}}{\mcP_h^{(s)}}=\frac{n}{2\sqrt{2}\dn} \frac{\cos^2\theta}{1+\cos^4\theta}\left(1-\frac{\dn}{n}\cos^2\theta\right).
\label{Phz}
\end{equation}
Thus they have the same order amplitudes.
\subsection{Detectability}
\label{Detectability}

In this subsection, we discuss the detectability of the sourced GWs in our model by CMB observations.
For the modes which exit the horizon during the attractor phase, $k\ge k_A$, the perturbations of the spectator fields never grow on super-horizon scales and thus the souring effects on GW perturbation as well as the curvature perturbation are not significant. Thus we focus on the CMB modes that exit the horizon during the growing phase, $k_{\rm CMB}\ll k_A$ in this subsection.

\begin{figure}[tbp]
    \hspace{-2mm}
  \includegraphics[width=70mm]{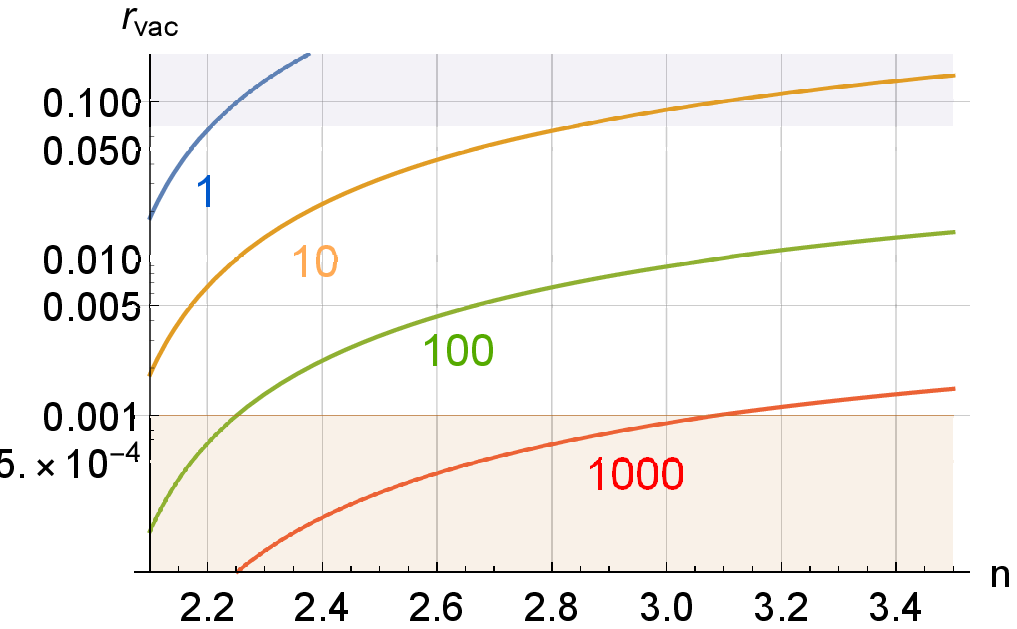}
  \hspace{5mm}
  \includegraphics[width=70mm]{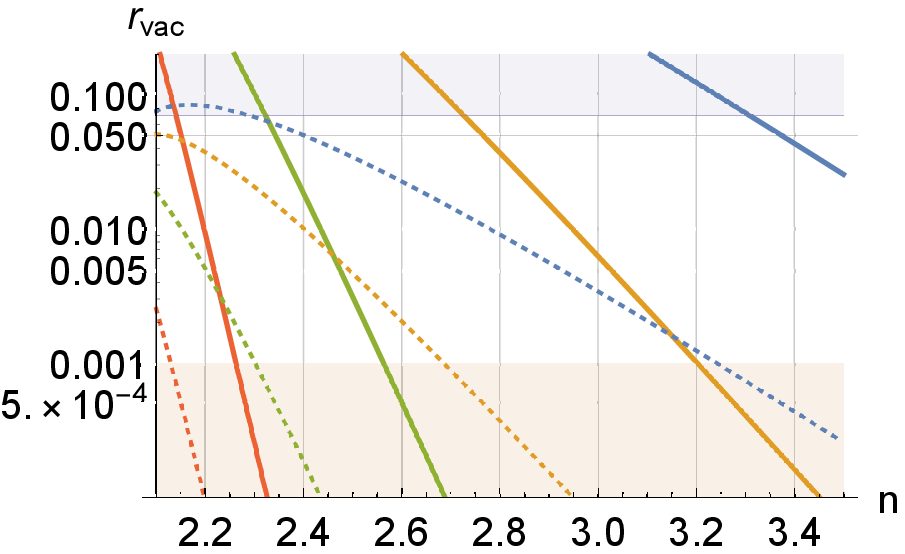}
  \caption
 {{\bf (Left panel)} The contour plot of $8\dn^2/(n^2\, r_{\rm vac})=1$ (blue), $10$ (yellow), $10^2$ (green) and $10^3$ (red). 
  $r_{\rm vac}>0.07$ and $r_{\rm vac}<10^{-3}$ are shaded in purple and orange, respectively. {\bf (Right panel)} The contour plot of the upper bound on $N_G$, eq.~\eqref{bound NG}. The solid lines show the bounds $N_G^{\max}=3$ (blue),
 $5$ (yellow), $10$ (green) and $20$ (red) for $\Lambda=10^{-2}\Mpl$. The dotted lines are for $\Lambda=10^{-3}\Mpl$ with the same colour scheme.}
 \label{R ratio}
\end{figure}

We consider the ratio between the vacuum contribution and the sourced one to the power spectra of the curvature  and GW perturbations,
\begin{align}
\mcR_\zeta \equiv \mcP_\zeta^{(s)}/\mcP_\zeta^{\rm (vac)},
\qquad\qquad
\mcR_h\equiv \mcP_h^{(s)}/\mcP_h^{\rm (vac)}
\end{align}
In order to have detectable sourced GWs without producing too large curvature perturbation, one needs to satisfy the following requirement,
\begin{equation}
\mcR_\zeta\ll 1, \qquad \mcR_h\gtrsim1.
\end{equation}
A necessary condition to satisfy them is 
\begin{equation}
\frac{\mcR_h}{\mcR_\zeta}=\frac{8\dn^2}{n^2r_{\rm vac}}\frac{1+\cos^4\theta}{\left(1-\frac{\dn}{n}\cos^2\theta\right)^2}\gg1,
\end{equation}
where we have used the so-called consistency\ relation, $r_{\rm vac}\equiv \mcP_h^{\rm (vac)}/\mcP_\zeta^{\rm (vac)}=16\epsilon_{\phi}$,
of single field slow-roll inflation.
For instance, if $\mcR_h/\mcR_\zeta$ is larger than 100, 
$\mcR_h>1$ and $\mcR_\zeta<10^{-2}$ are compatible.
In the left panel of figure~\ref{R ratio}, we plot the prefactor of $\mcR_h/\mcR_\zeta$, namely $8\dn^2/(n^2\, r_{\rm vac})$.
%
%

$\mcR_\zeta$ and $\mcR_h$ themselves contain three more parameters, namely $\Lambda,\ k_A/k$ and $N_A$ in addition to $n$ and $r_{\rm vac}$, and a systematic parameter survey is tricky. Nevertheless, it is not hard to find a favorable set of  parameters.
As an example, if one adopts the parameters,
\begin{equation}
n=2.5,\quad r_{\rm vac}=3\times 10^{-4},\quad
\Lambda=10^{-2}\Mpl,\quad k=e^{-5}k_A,
\quad N_A=15,
\label{ex para}
\end{equation}
one obtains
\begin{align}
\mcR_h &\approx 
21\left(1-\cos^2\theta+\cos^4\theta-\cos^6\theta\right),
\\
\mcR_\zeta &\approx
2\times 10^{-2}\left(1-1.4\cos^2\theta+0.44\cos^4\theta-0.04\cos^6\theta\right),\qquad
\end{align}
Therefore, in this case, the GW power spectrum is enhanced by 1 order of magnitude and becomes statistically anisotropic, while the correction to the curvature power spectrum is only $2\%$.
The maximum amplitude of the GW power spectrum at $\theta=\pi/2$
corresponds to 
\begin{equation}
r = 6.4\times 10^{-3},
\end{equation}
and it is detectable by upcoming CMB B-mode observations \cite{Matsumura:2013aja, Abazajian:2016yjj}.

Before closing this section, we discuss two constraints.
First, introducing 
\begin{equation}
N_G\equiv \ln[k_A/k],
\end{equation}
we put an upper bound on $N_G$. 
$N_G$ is the e-folding number from the horizon exit till the onset of the attractor phase, or the duration of the growing phase which the mode experiences,
\begin{equation}
\bar{\rho}_E(t_k) \exp[2\dn N_G]=  \frac{3}{2}\dn H^2 \Lambda^2
\quad\Longrightarrow\quad
N_G=\frac{1}{2\dn}\ln\left[\frac{3\dn H^2\Lambda^2}{2\bar{\rho}_E(t_k)}\right],
\end{equation}
where $t_k$ denotes the time when the $k$-mode of interest exits the horizon.
As $\bar{\rho}_E(t_k)$ is smaller, $N_G$ becomes larger.
However, for the validity of the perturbative approach $\bar{A}_i\gg \delta A_i$, $\bar{\rho}_E(t_k)$ should be much larger than $\mathcal{O}(H^4)$.
Requiring $\bar{\rho}_E(t_k)>10^2H^4$ and eliminating $H$
with $r_{\rm vac}=2H^2/(\pi^2\Mpl^2 \mcP_\zeta^{\rm obs})$,
we obtain the upper bound on $N_G$ as
\begin{equation}
N_G < N_G^{\max}\equiv\frac{1}{2\dn}
\ln\left[\frac{3\dn }{10^2\pi^2r_{\rm vac}\mcP_\zeta^{\rm obs}}\frac{\Lambda^2}{\Mpl^2}\right],
\label{bound NG}
\end{equation}
where $\mcP_\zeta^{\rm obs}\approx 2.2\times10^{-9}$.
This upper bound is plotted in the right panel of figure~\ref{R ratio}.
In the case of the parameters given in eq.~\eqref{ex para}, this upper bound leads to $N_G< 12.3$. 

We should also require that the energy density of the spectator scalar field is subdominant. Its energy fraction is given by
\begin{equation}
\Omega_\sigma \simeq \frac{V(\bar{\sigma})}{3\Mpl^2 H^2} = n \frac{\bar{\sigma} \Lambda}{\Mpl^2}.
\label{sigma fraction}
\end{equation}
Remembering $\dot{\bar \sigma}=-n\Lambda H$ during the growing phase and $\dot{\bar \sigma}=-2\Lambda H$ during the attractor phase which terminates at $\bar{\sigma}\simeq \Lambda$,
the field value of $\sigma$ can be estimated as
\begin{equation}
\bar{\sigma}(t_k)\simeq \left(nN_G+2N_A+1\right)\Lambda.
\end{equation}
Plugging it into eq.~\eqref{sigma fraction}, 
we obtain a constraint on the parameters,
\begin{equation}
\Omega_\sigma(t_k) \simeq n\left(nN_G+2N_A+1\right)
\frac{\Lambda^2}{\Mpl^2} \ll 1.
\end{equation}
In the case of the parameters given in eq.~\eqref{ex para}, 
$\Omega_\sigma(t_k)\approx 1.1\times 10^{-2}$ and the energy density of the spectator sector is subdominant. 

\section{Conclusion}
\label{Conclusion}

In this paper, we studied a model in which a spectator scalar field $\sigma$ is coupled to a $U(1)$ gauge field $A_\mu$, and their perturbations source the inflaton and GW perturbations during inflation. The background evolution of these spectator fields has the following three phases and their perturbations change the behaviors in each phase: (i) During the growing phase, $\bar{\rho}_E$ increases as $a^{2\dn}$ while its backreaction to $\bar{\sigma}$ is negligible. The intrinsic $\bar{I}\delta A^X_k/a$ and $\bar{I}\delta A_k^Y/a$ also grow as $a^{\dn}$ on super-horizon scales and $\delta\sigma_k$ sourced by $\delta A^X_k$ grows even faster as $a^{2\dn}$. (ii) As the backreaction becomes significant, the background dynamics of the spectator fields enters the attractor phase. Both the backgrounds and the super-horizon scale perturbations of the spectator fields stay constant.
Interestingly, we found that $\delta\sigma_k$ and $\delta A_k^X$ satisfy
the particular relation~\eqref{super horizon ratio}, which leads to the non-trivial anisotropies of the curvature and GW power spectra.
(iii) During the damping phase, $\bar{\sigma}$ starts damped oscillations and the gauge field energy density quickly decays as $a^{-4}$. Their perturbations also decay and only the sourced curvature perturbation and GWs remain as observables.
We have derived the analytic expressions for the background and the perturbations of the spectator fields and confirm them through numerical calculations.

The key predictions for the observables of this model are fivefold: 
(i) The sourced GW power spectrum has an interesting statistical anisotropy, eq.~\eqref{sourced Ph}, in which high multipole moments naturally appear.
(ii) The sourced GWs are linearly polarized and the respective polarizations have the different statistical anisotropies, eq.~\eqref{different SA}. (iii) The tensor-to-scalar ratio $r$ can be enhanced compared to the case of the  conventional vacuum fluctuation, with the GW power spectrum red-tilted, $\mcP_h^{(s)}\propto k^{-2\dn}$.
(iv) The sourced curvature perturbation can be much smaller than the observed curvature perturbation and the model can be consistent with the CMB constraints, although it also acquires a non-trivial statistical anisotropy, eq.~\eqref{sourced Pz}.
(v) The cross-correlation between the curvature  and GW perturbations,$\mcP_{h\zeta},$ is generated at the same level as $\mcP_h^{(s)}$ with a different statistical anisotropy, eq.~\eqref{Phz}.
We also discuss the parameter choices and the restrictions on the model in section~\ref{Detectability}.

If the power spectrum of the sourced GWs is larger than one-thousandth of the observed curvature perturbation on the CMB scale, the above predictions are potentially verifiable by the upcoming CMB B-mode observations such as LiteBIRD and CMB-S4.
Thus a forecast analysis based on these experiments would be an interesting future work.
It should be stressed that, however, we did not give a concrete mechanism for selecting the initial conditions to realize the growing phase, which is necessary to provide a sizable amount of GWs.
For the potential that maintains the condition $n\geq 2$, it might be natural that the background gauge field has already settled in the attractor value 
when the CMB scale modes cross the horizon.
Therefore we need to consider more complicated form of the potential $V(\sigma)$, in order for the proposed mechanism to work, while we exploit a toy potential eq.~\eqref{potential V} for simplicity in this work.
In light of its unique predictions, further studies on other potential forms based on a dedicated model building and the calculation of the power spectra in this model are also fascinating.
We leave them for future work.

\section*{Acknowledgement}

This work was supported in part by  MEXT KAKENHI Grant Numbers 17J09103 (T.F.), 15J01345 (I.O.), 17H06357 and 17H06358 (T.T.), and 15K17659 (S.Y.)
as well as the Grant-in-Aid for Scientific Research No. 26287044
and 15H02087 (T.T.), and 16H01103 (S.Y.).

\appendix

\section{Polarization Vector and Tensor}
\label{Polarization Vector and Tensor}

Here we  discuss the polarization vector and tensor.
The two linear polarization vectors whose wave number is parallel to the z-axis are written as%
\begin{equation}
\bm{e}^{X}(\hat{\bm z})= 
\begin{pmatrix}1 \\
0 \\
0 \\
\end{pmatrix},
\qquad
\bm{e}^{Y}(\hat{\bm z})= 
\begin{pmatrix}
0 \\
1 \\
0 \\
\end{pmatrix}
\end{equation}
To obtain the polarization vector with a general wave number $\hat{\bm k}$
which points the direction of $(\theta,\varphi)$ in polar coordinate, one uses the following rotation matrix
which transforms $\hat{\bm z}$ into $\hat{\bm k}$:
\begin{equation}
S(\hat{\bm k})=\begin{pmatrix}
\cos\theta \cos\varphi\ & -\sin\varphi\ & \sin\theta\cos\varphi \\
\cos\theta\sin\varphi & \cos\varphi & \sin\theta\sin\varphi \\
-\sin\theta & 0 & \cos\theta \\
\end{pmatrix}.
\end{equation}
Note that it is consistent with $\cos\theta= \bm{k}\cdot \dot{\bar{\bm A}}/(|\bm k| |\dot{\bar{\bm A}}|)$ and $\dot{\bar{\bm A}} \propto \hat{\bm z}$ (see \eqref{E prop z}).
Then one finds
\begin{equation}
\bm{e}^{X}(\hat{\bm k})=S(\hat{\bm k}) \bm{e}^{X}(\hat{\bm z})=
\begin{pmatrix}
\cos\theta\cos\varphi \\
\cos\theta\sin\varphi\\
-\sin\theta \\
\end{pmatrix},
\qquad
\bm{e}^{Y}(\hat{\bm k})=S(\hat{\bm k}) \bm{e}^{Y}(\hat{\bm z})=
\begin{pmatrix}
-\sin\varphi \\
\cos\varphi\\
0 \\
\end{pmatrix},
\label{Linear polarization vector}
\end{equation}
One can show that these polarization vectors satisfy
\begin{align}
&\bm{k}\cdot \bm{e}^{X/Y}(\hat{\bm k})=0,
\qquad
\bm{e}^{X}(-\hat{\bm k})=\bm{e}^{X}(\hat{\bm k}),
\qquad 
\bm{e}^{Y}(-\hat{\bm k})
=-\bm{e}^{Y}(\hat{\bm k}),\notag\\
&
\bm{e}^{X/Y}(\hat{\bm k})\cdot\bm{e}^{X/Y}(\hat{\bm k})=1,
\qquad
\bm{e}^{X/Y}(\hat{\bm k})\cdot\bm{e}^{Y/X}(\hat{\bm k})=0.
\label{pvector property}
\end{align}
With the linear polarization vectors, we define the following polarization tensor,
\begin{align}
e^+_{ij}(\hat{\bm k})&\equiv \frac{1}{\sqrt{2}}\left(
e^{X}_i(\hat{\bm k})e^{X}_j(\hat{\bm k})-e^{Y}_i(\hat{\bm k})e^{Y}_j(\hat{\bm k})\right),
\\
e^\times_{ij}(\hat{\bm k})&\equiv \frac{1}{\sqrt{2}}\left(
e^{X}_i(\hat{\bm k})e^{Y}_j(\hat{\bm k})+e^{Y}_i(\hat{\bm k})e^{X}_j(\hat{\bm k})\right).
\end{align}
These polarization tensors can be used as an orthonormal basis
of transverse-traceless tensors,
\begin{equation}
e^\lambda_{ij}(\hat{\bm k}) e^{\lambda'}_{ij}(\hat{\bm k})
=\delta^{\lambda\lambda'}.
\end{equation}
%

\section{Quadratic Action}
\label{Quadratic Action}

Here we show the full expression of the second order action
of $\dphi, \dsig, \delta A_i$ and $h_{ij}$,
\begin{equation}
S^{(2)}=\frac{1}{2}\int \dd t\dd^3 x\, a^3  \Big[ \mathcal{L}_{\rm scalar}
+\mathcal{L}_{\rm gauge} 
+\mathcal{L}_{\rm tensor}\Big].
\end{equation}
The Lagrangians of the scalar, gauge and tensor sectors are given by 
\begin{align}
\mathcal{L}_{\rm scalar}&= 
\dphid^2+\dsigd^2-a^{-2}(\partial_i \dphi)^2-a^{-2} (\partial_i \dsig)^2 -\mu_\phi^2 \dphi^2-\mu_\sigma^2\dsig^2-2\Omega_{\phi\sigma}\dphi\dsig,
\\
\mathcal{L}_{\rm gauge}&= 
\frac{\bar{I}^2}{a^2}\bigg[\delta \dot{A}_i^2-a^{-2}(\partial_i\delta A_j)^2
-\mu^{2}_{ij}\delta A_i \delta A_j -2 \Omega^{A\phi}_i  \delta A_i \dphi
- 2\Omega^{A\sigma}_i  \delta A_i \dsig
+4\frac{\bar{I}'}{\bar{I}} \dot{\bar{A}}_i\delta\dot{A}_i\dsig
\bigg],
\\
\mathcal{L}_{\rm tensor}&=  \frac{\Mpl^2}{4}\left(\dot{h}_{ij}\dot{h}_{ij}-a^{-2}\partial_k h_{ij} \partial_k h_{ij}\right) 
+\frac{\bar{I}^2}{a^2}h_{ik}h_{kj}\dot{\bar{A}}_i\dot{\bar{A}}_j
-\frac{2}{a^2} h_{ij} \left(\bar{I}^2\dot{\bar{A}}_i\delta\dot{A}_j+\bar{I} \bar{I}' \dot{\bar{A}}_i\dot{\bar{A}}_j \dsig \right)
\notag\\
&\quad+ \frac{\bar{I}^2 \dot{\bar{A}}_i \dot{\bar{A}}_j}{2a^2\Mpl^2 H} h_{ij}
\left(\dot{\bar \phi}\dphi +  \dot{\bar \sigma} \dsig+a^{-2}\bar{I}^2\dot{\bar{A}}_k\delta A_k \right),
\label{tensor L}
\end{align}
with
\begin{align}
\mu_\phi^2&\equiv 
\bar{U}''
-3\frac{\dot{\bar \phi}^2}{\Mpl^2}\left(1+\frac{\epsilon_H}{6}+\frac{2\ddot{\bar \phi}}{3H\dot{\bar \phi}}+\frac{\dot{\bar \phi}^2 +\dot{\bar \sigma}^2+2\bar{\rho}_E \sin^2\theta}{12\Mpl^2 H^2}\right),
\\
\mu_\sigma^2&\equiv 
\bar{V}''
+\frac{1}{a^2}\left(4{\bar I}'^2 \cos^2\theta -{\bar I}'^2-\bar{I} \bar{I}''\right)\dot{\bar{A}}_i^2
\notag\\
&\quad-3\frac{\dot{\bar \sigma}^2}{\Mpl^2}\left(1-\frac{2\bar{I} \bar{I}' \dot{\bar{A}}_i^2}{3a^2 H\dot{\bar \sigma}}\sin^2\theta+\frac{\epsilon_H}{6}+\frac{2\ddot{\bar \sigma}}{3H\dot{\sigma}}+\frac{\dot{\bar \phi}^2 +\dot{\bar \sigma}^2+2\bar{\rho}_E \sin^2\theta}{12\Mpl^2 H^2}\right),
\\
\Omega_{\phi\sigma} &\equiv 
-\frac{\dot{\bar \phi}\dot{\bar \sigma}}{\Mpl^2}\left[3-\frac{\bar{I} \bar{I}' \dot{\bar{A}}_i^2}{a^2 H\dot{\bar \sigma}}\sin^2\theta+\frac{\epsilon_H}{2}+
\frac{\dot{\bar \phi}^2 +\dot{\bar \sigma}^2+\bar{\rho}_E\sin^2\theta}{4\Mpl^2 H^2}+\frac{\ddot{\bar \phi}}{H\dot{\bar \phi}}+\frac{\ddot{\bar \sigma}}{H\dot{\bar \sigma}}\right],
\\
\mu^{2}_{ij}&\equiv \frac{3\bar{I}^2 \dot{\bar{A}}_i \dot{\bar{A}}_j}{a^2\Mpl^2}\left(1-\frac{\epsilon_H}{6}-\frac{\dot{\bar \phi}^2+\dot{\bar \sigma}^2+2\bar{\rho}_E\sin^2\theta}{12\Mpl^2H^2}\right),
\\
\Omega^{A\phi}_i &\equiv -\frac{\dot{\bar \phi}\dot{\bar{A}}_i}{2\Mpl^2}\left(\epsilon_H+\frac{2\ddot{\bar \phi}}{H\dot{\bar \phi}}+\frac{\dot{\bar \phi}^2+\dot{\bar \sigma}^2+2\bar{\rho}_E \sin^2\theta}{2\Mpl^2 H^2}\right),
\\
\Omega^{A\sigma}_i &\equiv  \frac{\dot{\bar{A}}_i}{2\Mpl^2}\left[2\frac{\bar{I} \bar{I}' \dot{\bar{A}}_j^2}{a^2 H}\sin^2\theta-\dot{\bar \sigma}\left(\epsilon_H+\frac{2\ddot{\bar \sigma}}{H\dot{\bar \sigma}}+\frac{\dot{\bar \phi}^2+\dot{\bar \sigma}^2+2\bar{\rho}_E \sin^2\theta}{2\Mpl^2 H^2}\right)\right],
\end{align}
where $\epsilon_H\equiv -\dot{H}/H^2$.

\section{Super-horizon solutions during attractor phase}
\label{Super-horizon solutions during attractor phase}

We solve the coupled equation of $\delta\sigma_k$ and $\delta A^X_k$. It is useful to rewrite the EoMs with the cosmic time as
\begin{align}
\ddot{\delta\sigma}_k+3H\dot{\delta\sigma}_k+\left(\frac{k^2}{a^2}+\mu_\sigma^2\right)\delta\sigma_k
&=-2\sqrt{2}\frac{\sqrt{\bar{\rho}_E}}{\Lambda}\sin\theta\ \frac{\bar{I}\delta\dot{A}_k^X}{a},
\label{dsig EoM t}
\\
\partial_t \left(\frac{\bar{I}\delta\dot{A}_k^X}{a}\right)+
a^{-3}\partial_t \left(a^2\bar{I}\right) \delta \dot{A}_k^X
+\frac{k^2}{a^2}\delta A_k^X  &=
2\sqrt{2}\frac{\sqrt{\bar{\rho}_E}}{\Lambda}\sin\theta\ \delta\dot{\sigma}_k.
\label{dA EoM t}
\end{align}
During the attractor phase, we can use
\begin{equation}
\mu_\sigma^2=6\dn \cos(2\theta)H^2,
\quad
\frac{\sqrt{\bar{\rho}_E}}{\Lambda}=
\sqrt{\frac{3}{2}\dn} H,
\quad \partial_t \left(a^2\bar{I}\right) =
a^2\bar{I}\left(2H+\frac{\dot{\bar \sigma}}{\Lambda}\right)=0.
\end{equation}
Then eqs.~\eqref{dsig EoM t} and \eqref{dA EoM t} read
\begin{align}
\ddot{\delta\sigma}_k+3H\dot{\delta\sigma}_k+\left(\frac{k^2}{a^2}+6\Delta n\cos(2\theta)H^2\right)\delta\sigma_k
&=-2
\sqrt{3\Delta n}  \sin\theta\ H\frac{\bar{I}\delta\dot{A}_k^X}{a},
\\
\partial_t \left(\frac{\bar{I}\delta\dot{A}_k^X}{a}\right)+\frac{k^2}{a^3}\bar{I}\delta A_k^X  &=2
\sqrt{3\Delta n} \sin\theta\ H \delta\dot{\sigma}_k.
\end{align}
On super-horizon scale, we can ignore the gradient terms (i.e. the terms with $k^2$) and the second equation is solved as
\begin{equation}
\frac{\bar{I}\delta\dot{A}_k^X}{a}=
2\sqrt{3\Delta n} \sin\theta\ H \delta\sigma_k+
D_1,
\qquad {\rm (super\ horizon)}
\end{equation}
where $D_1$ is the integration constant.
Substituting it into the first equation, we find
\begin{equation}
\ddot{\delta\sigma}_k+3H\dot{\delta\sigma}_k+6\Delta nH^2\delta\sigma_k
=
\tilde{D}_1,
\qquad {\rm (super\ horizon)}
\end{equation}
where $\tilde{D}_1\equiv-2\sqrt{3\Delta n}  \sin\theta\ H D_1$.
Its solution is given by
\begin{align}
\delta\sigma_k=&C_1\exp\left[\frac{1}{2}\left(\sqrt{9-24\Delta n}-3\right)H(t-t_k)\right]
+ C_2 \exp\left[-\frac{1}{2}\left(\sqrt{9-24\Delta n}+3\right)H(t-t_k)\right]
\notag\\
&-\frac{D_{1}}{\sqrt{3\Delta n}H}\sin\theta,
\qquad {\rm (super\ horizon)}
\label{delta sigma C1C2}
\end{align}
where $C_1, C_2$ are the integration constant and $t_k$ is a certain time when the super horizon approximation becomes good.
Since the first and second term decay, the third term becomes dominant. We also obtain the solution of the gauge field perturbation,
\begin{align}
\frac{\bar{I}(t_k)}{a(t_k)}\delta A_k^X &=\frac{D_1}{3H}\cos(2\theta)\, e^{3H(t-t_k)}  +D_2 
\notag\\
 &+ \frac{\sin\theta}{2\sqrt{\Delta n}}  \bigg\{C_1\left(\sqrt{3}-\sqrt{3-8\Delta n}\right) e^{\frac{1}{2}\left(3+\sqrt{9-24\Delta n}\right)H(t-t_k)}
\notag\\&~~~~~~~~~~~~~~~+ C_2 \left(\sqrt{3}+\sqrt{3-8\Delta n}\right) e^{\frac{1}{2}\left(3-\sqrt{9-24\Delta n}\right)H(t-t_k)}\bigg\},
\quad {\rm (super\ horizon)}
\label{delta A D1D2}
\end{align}
The first term eventually becomes dominant.
Comparing these leading terms, we obtain eq.~\eqref{super horizon ratio}.

\end{document}